\documentclass[twocolumn,prl,showpacs,preprintnumbers,amsmath,amssymb]{revtex4}
\usepackage{graphicx}
\usepackage{dcolumn}
\usepackage{bm}

\begin{document}

\title{Strain modulation of transport criticality in heterogeneous solids}
\author{$^1$Sonia Vionnet, $^1$Claudio Grimaldi, $^{1,2}$Thomas Maeder,
$^{1,2}$Sigfrid Str\"assler, and $^1$Peter Ryser, }
\affiliation{$^1$ Laboratoire de Production Microtechnique, Ecole
Polytechnique F\'ed\'erale de Lausanne, CH-1015 Lausanne,
Switzerland} \affiliation{$^2$Sensile Technologies SA, PSE,
CH-1015 Lausanne, Switzerland}

\begin{abstract}
A vast class of disordered conducting-insulating compounds close
to the percolation threshold is characterized by nonuniversal
values of transport critical exponents. The lack of universality
implies that critical indexes may depend on material properties
such as the particular microstructure or the nature of the
constituents, and that in principle they can be influenced by
suitable applied perturbations leading to important informations
about the origin of nonuniversality. Here we show that in
RuO$_2$-glass composites the nonuniversal exponent can be
modulated by an applied mechanical strain, signaled by a
logarithmic divergence of the piezoresistive response at the
percolation threshold. We interpret this phenomenon as being due
to a tunneling-distance dependence of the transport exponent,
supporting therefore a theory of transport nonuniversality
proposed some years ago.
\end{abstract}

\maketitle

Transport properties of disordered insulator-conductor composites
are characterized by the existence of a percolation critical
value $x_c$ of the conducting phase volume concentration $x$
below which the system becomes an insulator. As $x-x_c\rightarrow
0$ the resistivity $\rho$ of the composite diverges by following
a power-law behavior of the form:
\begin{equation}\label{rho}
\rho\simeq \rho_0 (x-x_c)^{-t},
\end{equation}
where $\rho_0$ is a material-dependent prefactor and $t$ is the
transport critical exponent\cite{stauffer,sahimi}. According to
the standard theory of transport in isotropic percolating
materials, $\rho_0$ and $x_c$ depend on microscopic details such
as the microstructure and the mean inter-grain junction
resistivity while, on the contrary, the exponent $t$ is material
independent and takes on the value $t\simeq 2.0$ for three
dimensional systems\cite{stauffer,sahimi}. This universal value
is indeed observed in a great variety of
composites\cite{abeles,lee,chen,dziedzic} and it is confirmed to a
rather high accuracy by calculations on random resistor networks
models\cite{batrouni}.

Universality of DC transport is understood as being due to the
unimportance of microscopic details for the flow of current
compared to the topological properties of the percolating
backbone\cite{stauffer}. This concept applies also to other
transport-percolation phenomena such as for example proton
conductivity in biological materials and laminar fluid flows
through porous media\cite{sahimi2}. Yet, a vast class of
disordered composites deviates from universality by displaying
values of $t$ larger than $2.0$\cite{dziedzic,wu} and as high as
$t\simeq 7.0$\cite{pike} or even more\cite{soares}.  The lack of
universality indicates that the exponent $t$ acquires an
additional dependence upon some microscopic properties and
variables, whose identification would permit to unveil the
mechanisms leading to $t\neq 2.0$. A practical route to achieve
this goal is to apply an external perturbation to a nonuniversal
composite with the hope of affecting the same microscopic
variables which govern nonuniversality. In this case, the applied
perturbation would change the value of the critical exponent $t$.

We have applied this idea to RuO$_2$-glass composites prepared as
thick-film resistors (TFRs) in our laboratory. RuO$_2$ as well as
other ruthenate (Pb$_2$Ru$_2$O$_6$, Bi$_2$Ru$_2$O$_7$) or oxide
(IrO$_2$) TFRs are ideal systems for our study because they
display universal ($t\simeq 2.0$) or nonuniversal
($t>2.0$\cite{dziedzic} up to about $t\simeq 7.0$\cite{pike})
behavior of transport depending on the microstructure and
fabrication procedures, with a good reproducibility and high
stability. In these composites, the conducting particles are
separated by nanometer-thick film of glass\cite{chiang} and the
dominant mechanism of transport is via quantum tunneling through
the glass films\cite{chiang,pike2,prude}. Here we show that when
transport is nonuniversal, the critical exponents of our samples
can be modified by an applied mechanical strain $\varepsilon$,
providing evidence that the microscopic variable governing
nonuniversality is the mean tunneling distance $a$ between
adjacent conducting grains.

Our samples were prepared by mixing two series of RuO$_2$ powders
with 40 nm and 400 nm grain sizes, respectively, with a
lead-borosilicate glass powder [PbO(75\% wt)-B$_2$O$_3$(10\%
wt)-SiO$_2$ (15\% wt)] of $1-5$ $\mu$m grain sizes together with a
vehicle of terpineol and ethyl cellulose. $2\%$ of Al$_2$O$_3$
were added to the pastes in order to avoid crystallization. The
pastes were screen printed on Al$_2$O$_3$ substrates with gold
electrical contacts and fired for $15$ minutes at temperatures
$T_f$ (see Table \ref{table1}) higher then the temperature of
fusion of the glass (about $500^\circ$C). The resulting films
were about 10 $\mu$m thick and several resistivity measurements
were taken over eight different samples for each RuO$_2$ volume
fraction value.

\begin{figure}[t]
\protect
\includegraphics[scale=0.42]{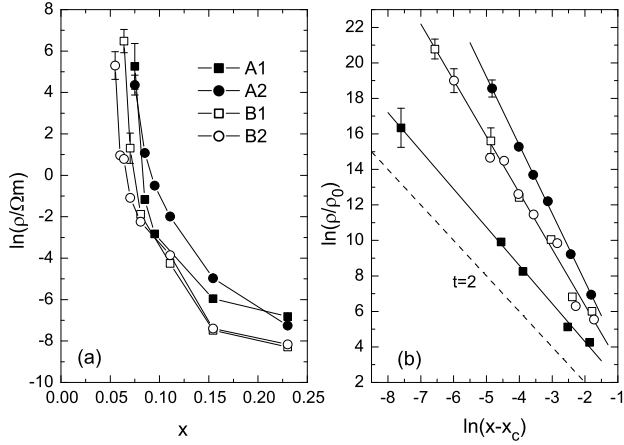}
\caption{(a): resistivity $\rho$ as a function of RuO$_2$ volume
concentration $x$ for four different series of TFRs. The solid
lines are a guide to the eyes. (b) ln-ln plot of the same data of
(a) with fits to Eq.(\ref{rho}) shown by solid lines. The dashed
line has slope $t=2$ corresponding to universal behavior of
transport. The prefactor $\rho_0$, critical concentration $x_c$
and transport exponent $t$ values obtained by the fits are
reported in Table \ref{table1}} \label{fig1}
\end{figure}

In Fig.\ref{fig1} we report the room temperature resistivity
$\rho$ measured for four different series of TFRs (see Table
\ref{table1}) as functions of the RuO$_2$ volume concentration
$x$. As shown in Fig.\ref{fig1}(a), resistivity diverges at rather
small values of $x$, as expected when the mean grain size of the
conducting phase (40 nm and 400 nm) is smaller than that of the
glass (1-5 $\mu$m)\cite{kusy}. The same data are re-plotted in
the ln-ln plot of Fig.\ref{fig1}(b) together with the
corresponding fits to Eq.(\ref{rho}) (solid lines) and the
best-fit parameters $\rho_0$, $x_c$ and $t$ are reported in Table
\ref{table1} . As it is clearly shown, our resistivity data follow
the power law behavior of Eq.(\ref{rho}) with exponent $t$ close
to the universal value $t\simeq 2.0$ for the A1 series ($t=2.15\pm
0.06$) or markedly nonuniversal as for the A2 case which displays
$t=3.84\pm 0.06$. The B1 and B2 series have virtually equal
values of $t$ ($t\simeq 3.16$) falling in between those of the A1
and A2 series.

\begin{table*}
\caption{\label{table1}Label legend of the various sample used in
this work with fitting parameters of Eqs.(\ref{rho},\ref{gamma2})
}
\begin{ruledtabular}
\begin{tabular}{cccccccccc}
Label & RuO$_2$ grain size & firing temperature $T_f$ & $x_c$ &
$\ln(\rho_0/\Omega{\rm m})$ & $t$
 & $\Gamma_0$ & $dt/d\varepsilon$  \\
\hline A1 & 400nm & $525^\circ$C  & $0.0745$ & $-11.1\pm 0.3$ &
$2.15\pm 0.06$ & $5.5\pm 1.5$ & $-0.2\pm 0.4$  \\
A2 & 400nm & $600^\circ$C  & $0.0670$ & $-14.2\pm 0.2$ & $3.84\pm
0.06$ & $-8.8\pm 1.6$ & $5.4\pm 0.5$  \\
B1 & 40nm & $550^\circ$C & $0.0626$ & $-14.3\pm 0.5$ & $3.17\pm
0.16$ & $-15.3\pm 3$ & $8.7\pm 0.9$  \\
B2 & 40nm & $600^\circ$C & $0.0525$ & $-13.7\pm 0.7$ & $3.15\pm
0.17$ & $-19.3\pm 2.4$ & $11.0\pm 0.7$   \\
\end{tabular}
\end{ruledtabular}
\end{table*}

The effect of an applied strain $\varepsilon$ on transport is
obtained by recording the piezoresistive response, {\it i. e.}
the relative resistivity change $\Delta \rho/\rho$ upon applied
mechanical strain deformation, by cantilever bar measurements.
The RuO$_2$-glass pastes were screen printed on Al$_2$O$_3$
cantilever bars 51 mm long, $b=5$ mm large, and $h=0.63$ mm
thick. The cantilever was clamped at one end and different
weights were applied at the opposite end. The resulting substrate
strain $\varepsilon$ along the main cantilever axis can be
deduced from the relation $\varepsilon=6Mgd/(Eb h^2)$, where
$d=27.8$ mm is the distance between the resistor and the point of
applied force, $E=332.6$ GPa is the reduced Al$_2$O$_3$ Young
modulus, $g$ is the gravitational acceleration and $M$ is the
value of the applied weight. As shown in
Refs.\onlinecite{grima2,grima3}, from measurements of the
longitudinal ($\Delta\rho_\parallel/\rho$) and transverse
($\Delta\rho_\perp/\rho$) piezoresistive responses obtained by
measuring the voltage drops along and perpendicular to the main
cantilever axis, respectively, it is possible to extract the
isotropic resistive variation defined as $\Delta\rho/\rho=(\Delta
\rho_\parallel/\rho+2\Delta\rho_\perp/\rho)/3$. For the values of
RuO$_2$ content used in this work,
$\Delta\rho_\parallel/\rho\simeq \Delta\rho_\perp/\rho$
confirming the vicinity to a percolation
threshold\cite{grima2,grima3}.

\begin{figure}
\protect
\includegraphics[scale=0.42]{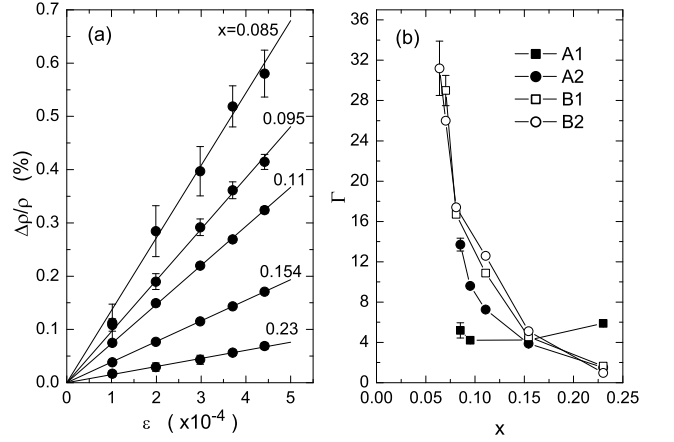}
\caption{(a) relative variation of resistivity as a function of
applied strain $\varepsilon$ in cantilever bar measurements of
the A2 series for different contents $x$ of RuO$_2$. The solid
lines are linear fits to the data. (b) isotropic piezoresistivity
response $\Gamma=d\ln(\rho)/d\varepsilon$ obtained from the
slopes of (a) as a function of the volume concentration $x$ for
the TFRs series of Fig.\ref{fig1}.} \label{fig2}
\end{figure}

In Fig.\ref{fig2}(a) we report $\Delta \rho/\rho$ for various
RuO$_2$ concentrations $x$ of the A2 series as a function of
applied strain $\varepsilon$. In the whole range of applied
strains, $\Delta \rho/\rho$ changes linearly with $\varepsilon$
permitting to extract rather accurate piezoresistive factors
$\Gamma=d\ln(\rho)/d\varepsilon$ from the slopes of the linear
fits of $\Delta \rho/\rho$ {\it vs} $\varepsilon$. The so obtained
$\Gamma$ values as a function of $x$ are plotted in
Fig.\ref{fig2}(b) for the four TFRs series of Fig.\ref{fig1}.
With the exception of the A1 series, $\Gamma$ displays a strong
dependence upon $x$ and tends to diverge as $x$ approaches to the
same critical concentrations $x_c$ at which $\rho$ goes to
infinity.

The divergence of $\Gamma$ at $x_c$ has been reported some years
ago\cite{carcia}, but here we show that it is caused by a strain
modulation of the critical exponent $t$. To see this, let us
consider Eq.(\ref{rho}) and differentiate it with respect to
$\varepsilon$:
\begin{equation}
\label{gamma1} \frac{d\ln(\rho)}{d\varepsilon}=\Gamma=\Gamma_0+
\frac{d}{d\varepsilon}\left[t\ln\left(\frac{1}{x-x_c}\right)\right].
\end{equation}
Since $\Gamma_0=d\ln(\rho_0)/d\varepsilon$ is a constant
independent of $x$, all the $x$ dependence of $\Gamma$ must come
from the last term of Eq.(\ref{gamma1}). However, it must be
noted that $x$ is a measure of the concentration $p$ of intergrain
junctions with finite resistances present in the
sample\cite{stauffer,sahimi}. In our measurements, the applied
strain is only of order $10^{-4}$, which is so small that it can
eventually change the value of the tunneling junction resistances
but cannot modify the concentration $p$ of junctions. So
$dx/d\varepsilon$ must be zero and by the same token also
$dx_c/d\varepsilon=0$. Hence the only way to have a $x$
dependence of $\Gamma$ is to allow the transport exponent $t$ to
have non vanishing derivative, which reduces Eq.(\ref{gamma1}) to:
\begin{equation}
\label{gamma2}
\Gamma=\Gamma_0+\frac{dt}{d\varepsilon}\ln\left(\frac{1}{x-x_c}\right).
\end{equation}

\begin{figure}[b]
\protect
\includegraphics[scale=0.62]{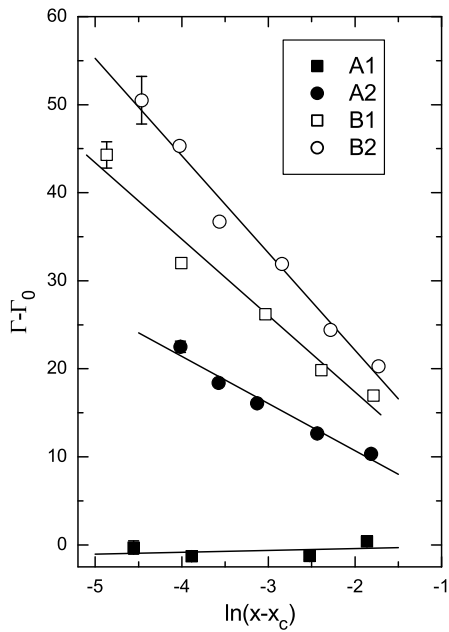}
\caption{Piezoresistive factor as a function of $\ln(x-x_c)$ and
fits (solid lines) to Eq.(\ref{gamma2}). The fit parameters
$dt/d\varepsilon$ and $\Gamma_0$ are reported in Table I. Only
sample A1 (solid squares) has quasi-zero slope indicating that
$dt/d\varepsilon=0$.} \label{fig3}
\end{figure}

In Fig.\ref{fig3} we plot the piezoresistive data of
Fig.\ref{fig2}(b) as a function of $\ln(x-x_c)$ with the same
values of the critical concentrations $x_c$ extracted from the
resistivity data. Our data fit well with Eq.(\ref{gamma2}) (solid
lines and last two columns of Table I) and indicate that $\Gamma$
indeed diverges logarithmically as $x\rightarrow x_c$ for the
series A2, B1, and B2. An analysis of previously published
data\cite{carcia} confirms this conclusion\cite{grima1}. It is
also remarkable that the series A1 has vanishing
$dt/d\varepsilon$ within experimental errors, in perfect
agreement with the universal value of its exponent ($t\simeq
2.15\pm 0.06$). We conclude therefore that when the transport
exponents of our RuO$_2$ TFRs are nonuniversal, they can be
modulated by an applied strain.

Let us discuss now what this finding implies in terms of the
microscopic origin of nonuniversality. The main effect of the
applied strain on the microscopic properties of RuO$_2$-glass
composites is that of changing the mean intergrain tunneling
distance $a$, $a\rightarrow a(1+\varepsilon)$, leading to a
variation of the microscopic tunneling resistances. Hence, in
order to have $dt/d\varepsilon\neq 0$, the transport exponent
itself must depend on $a$. A scenario of this kind has been
proposed by Balberg a few years ago in his tunneling-percolation
theory of nonuniversality in carbon-black--polymer
composites\cite{balb}. According to this theory, when the
distribution function of the tunneling distance $d$ between two
neighboring grains decays with $d$ much slower than the tunneling
decay $\exp(-2d/\xi)$, where $\xi$ is the localization length,
then the distribution function $h(r)$ of the intergrain tunneling
resistances $r$ develops a power-law tail such that $h(r)\propto
r^{\alpha-2}$ for $r\rightarrow \infty$ where
$\alpha=1-\xi/(2a)$. It is well known\cite{kogut} that
distribution functions of this kind lead to an explicit
dependence of $t$ upon the parameter $\alpha$\cite{stenull,alava}:
\begin{equation}
\label{nonuni} t=\left\{
\begin{array}{ll}
t_0 & \mbox{if} \hspace{3mm} \nu+\frac{1}{1-\alpha}< t_0 \\
\nu+\frac{1}{1-\alpha} &  \mbox{if}\hspace{3mm}
\nu+\frac{1}{1-\alpha}> t_0
\end{array}
\right.,
\end{equation}
where $t_0\simeq 2.0$ is the universal transport exponent and
$\nu\simeq 0.88$ is the 3D correlation-length exponent. In view
of Eq.(\ref{nonuni}), the term which multiplies the logarithm in
Eq.(\ref{gamma2}) is $dt/d\varepsilon=1/(1-\alpha)=2a/\xi > 0$
when $t$ is given by the second line of Eq.(\ref{nonuni}) or
$dt/d\varepsilon=0$ when $t=t_0$. However, we note that in
RuO$_2$-based TFRs the large difference between the bulk modulus
$B$ values of RuO$_2$ and of the glass ($B_{{\rm RuO}_2}\simeq
270$GPa and $B_{\rm glass}\simeq 40-80$GPa) leads to local strain
variations resulting basically in a strong $\varepsilon$
amplification within the softer phase (glass) through which
tunneling occurs. Hence, the $dt/d\varepsilon$ values of A2, B1
and B2 reported in Table I are not simply equal to $2a/\xi$ but
incorporate also the amplification effect of the strain
heterogeneity.

An interesting feature of our piezoresistive data is the change of
sign of $\Gamma_0$ which from positive for the universal serie A1
becomes negative for the nonuniversal ones (A2,B1 and B2, see
Table \ref{table1}). Although being quite surprising at first
sight, this result can be naturally explained within the
tunneling-percolation model\cite{balb}. In fact a tensile strain
($\varepsilon>0$) enhances the intergrain tunneling resistances
leading to an overall enhancement of the sample resistivity, so
that $\Gamma=d\ln(\rho)/d\varepsilon$ must be positive. Hence,
from Eq.(\ref{gamma2}), $\Gamma_0>0$ when $dt/d\varepsilon=0$
while when $dt/d\varepsilon>0$ $\Gamma_0$ does not need to be
positive to ensure $\Gamma>0$. Quite remarkably, an effective
medium calculation with the tunneling-percolation distribution
function $h(r)\propto r^{\alpha-2}$ leads to $\Gamma_0<0$ as long
as $t$ depends upon $\alpha$ as in the second line of
Eq.(\ref{nonuni})\cite{grima1}. This is in agreement with our
experimental results (see Table I).

Let us now turn to discuss briefly how other theories of
nonuniversality fit our findings. First, the random void model
(RVM) of continuum percolation\cite{halpe} can hardly be applied
to the A1 and A2 series, since the RuO$_2$ particles are not
microscopically smaller than the glass grains, so that the
conducting phase cannot be regarded as a continuum. In addition,
the RVM predicts that $t$ follows Eq.(\ref{nonuni}) with
$\alpha=\alpha_{\rm RVM}$ being a purely topological quantity
defined by the distribution of the narrow necks bounded by three
interpenetrating insulating spheres. Hence, $\alpha_{\rm RVM}$ is
insensitive to an applied strain so that $dt/d\varepsilon=0$ and
the piezoresistivity $\Gamma$ cannot develop a logarithmic
divergence. We exclude also the possibility that our nonuniversal
samples are in the mean-field regime for which $t=t_{\rm
mf}=3$\cite{heaney}. In fact the A2 series has a critical
exponent clearly larger than $t_{\rm mf}$ and, more importantly,
$dt_{\rm mf}/d\varepsilon$ is expected to be strictly zero.

In summary, we have shown that the piezoresistive response of
disordered RuO$_2$-glass composites has a logarithmic divergence
at the percolation threshold when DC transport is nonuniversal. A
coherent interpretation of this result calls into play a mean
tunneling distance dependence of the resistivity exponent, in
agreement with a tunneling-percolation origin of nonuniversality
proposed some time ago\cite{balb}. Such mechanism of universality
breakdown could apply also to other materials for which transport
is governed by tunneling such as carbon-black--polymer composites,
and experiments on their piezoresistive response could confirm
such conjecture.

Before concluding, it is worth to point out that in addition to
the importance of our results concerning the origin of transport
nonuniversality in tunneling-percolation materials, the reported
logarithmic divergence of $\Gamma$ has also consequences for
applications of Ru-based TFRS in force and pressure sensors. In
fact, the efficiency of piezoresistive sensors is measured by the
magnitude of $\Gamma$. However the logarithmic increase of
$\Gamma$ as $x\rightarrow x_c$ is not sufficient to overcome the
power-law divergence of the resistance fluctuations. This poses
serious limitations regarding the precision of pressure/force
detection. An alternative route would be to make use of the
amplification effects induced by the local elastic
heterogeneities by fabricating RuO$_2$-based TFRs with softer
insulating phases like low melting temperature glasses or even
polymeric hosts.

This work was partially supported by TOPNANO 21 (project NAMESA,
No. 5557.2).

\end{document}